\documentclass[11pt]{article}
\usepackage{graphicx}
\usepackage[margin=1.25in]{geometry}
\usepackage[usenames,dvipsnames]{color}
\usepackage{url}
\usepackage[colorlinks = true,
            linkcolor = blue,
            urlcolor  = blue,
            citecolor = blue,
            anchorcolor = blue]{hyperref}
\usepackage{lineno}

\bigskip
\bigskip

\def\Title#1{\begin{center} {\Large #1 } \end{center}}
\def\Author#1{\begin{center}{ \sc #1} \end{center}}
\def\Address#1{\begin{center}{ \it #1} \end{center}}

\newenvironment{Abstract}{\begin{quotation} \begin{center}
                       ABSTRACT
     \end{center}\bigskip  }{\end{quotation}}

\def\beq{\begin{equation}}
\def\eeq#1{\label{#1}\end{equation}}
\def\eeqn{\end{equation}}

\newenvironment{Eqnarray}%
   {\arraycolsep 0.14em\begin{eqnarray}}{\end{eqnarray}}
\def\beqa{\begin{Eqnarray}}
\def\eeqa#1{\label{#1}\end{Eqnarray}}
\def\eeqan{\end{Eqnarray}}

\let\bar=\overbar

\def\lsim{\mathrel{\raise.3ex\hbox{$<$\kern-.75em\lower1ex\hbox{$\sim$}}}}
\def\gsim{\mathrel{\raise.3ex\hbox{$>$\kern-.75em\lower1ex\hbox{$\sim$}}}}

\def\del{\partial}
\def\Dslash{\not{\hbox{\kern-4pt $D$}}}
\def\dslash{\not{\hbox{\kern-2pt $\del$}}}
\def\pslash{\not{\hbox{\kern-2pt $p$}}}
\def\ETmiss{\not{\hbox{\kern-4pt $E$}}_T}

\def\Dlr{\mathrel{\raise1.5ex\hbox{$\leftrightarrow$\kern-1em\lower1.5ex\hbox{$D$}}}}

\def\MSB{{\bar{M \kern -2pt S}}}
\def\msb{{\bar{\scriptsize M \kern -1pt S}}}

\def\drb{{\bar{\scriptsize D \kern -1pt R}}}

\begin{document}


\Title{\bf{Novel Sensors for Particle Tracking: A Contribution to the Snowmass Community Planning Exercise of 2021}}
\medskip
\Author{M.R. Hoeferkamp, S. Seidel\footnote{contact: seidel@unm.edu}}
\Address{Department of Physics and Astronomy, University of New Mexico, Albuquerque, NM, USA}
\Author{S. Kim, J. Metcalfe, A. Sumant}
\Address{Physics Division, Argonne National Laboratory, Lemont, IL, USA}
\Author{H. Kagan}
\Address{Department of Physics, Ohio State University, Columbus, OH, USA}
\Author{W. Trischuk}
\Address{Department of Physics, University of Toronto, Toronto, ON, Canada}
\Author{M. Boscardin}
\Address{Fondazione Bruno Kessler, Trento, Italy}
\Author{G.-F. Dalla Betta}
\Address{Department of Industrial Engineering, University of Trento, Trento, Italy}
\Author{D.M.S. Sultan}
\Address{Trento Institute for Fundamental Physics and Applications, INFN Trento, Trento, Italy}
\Author{N.T. Fourches}
\Address{CEA-Saclay, Universit\'e Paris-Saclay, Paris, France}
\Author{C. Renard}
\Address{CNRS-C2N, Universit\'e Paris-Saclay, Paris, France}
\Author{A. Barbier}
\Address{CEA-Iramis, Universit\'e Paris-Saclay, Paris, France}
\Author{T. Mahajan, A. Minns, V. Tokranov, M. Yakimov, S. Oktyabrsky}
\Address{SUNY College of Nanoscale Science and Engineering, Albany, NY, USA}
\Author{C. Gingu, P. Murat}
\Address{Fermi National Accelerator Laboratory, Batavia, IL, USA}
\Author{M.T. Hedges}
\Address{Purdue University, West Lafayette, IN, USA}

 \begin{Abstract}
\noindent Five contemporary technologies are discussed in the context of their potential roles in particle tracking for future high energy physics applications.  These include sensors of the 3D configuration, in both diamond and silicon, submicron-dimension pixels, thin film detectors, and scintillating quantum dots in gallium arsenide. Drivers of the technologies include radiation hardness, excellent position, vertex, and timing resolution, simplified integration, and optimized power, cost, and material.
\end{Abstract}
\clearpage

\def\thefootnote{\fnsymbol{footnote}}
\setcounter{footnote}{0}
\begin{center}
\section*{Executive Summary}
\end{center}

Five contemporary technologies with potential
application to particle tracking in future high energy physics experiments are discussed.

\bigskip
Silicon sensors of the 3D technology have electrodes oriented perpendicular to their wafer
surfaces.  These show promise for compensation of lost signal in high radiation
environments and for separation of pileup events by precision timing.  New 3D geometries involving p-type trench electrodes spanning the entire length of the
detector, separated by lines of segmented n-type electrodes for readout, promise improved uniformity, timing resolution, and radiation resistance relative to
established devices operating effectively at the LHC.  Present research
aims for operation with adequate
signal-to-noise ratio at fluences approaching $10^{18}~n_{\rm eq}/{\rm cm}^2$, with timing resolution
on the order of 10 ps.

\bigskip
The 3D technology is also
being realized in diamond substrates, where column-like electrodes may be placed
inside the detector material by use of a 130 fs laser with wavelength 800~nm.  When
focussed to a 2 micron spot, the laser has energy density sufficient to convert diamond
into an electrically resistive mixture of different carbon phases.  The drift
distance an electron-hole pair must travel to reach an electrode can be reduced
below the mean free path without reducing the number of pairs created.  Initial
tests have shown that after $3.5 \times 10^{15}~n_{\rm eq}/{\rm cm}^2$, a 3D diamond sensor with $50~\mu {\rm m} \times 50~\mu {\rm m}$ cells collects more charge than would be collected by a
planar device and shows less damage due to the shorter drift distance. 
The goal
of this project is to create a detector that is essentially immune to radiation
doses at the level of $10^{17}$ hadrons/cm$^2$.

\bigskip
A pixel architecture, named DoTPiX, has been proposed on the principle of a
single n-channel MOS transistor, in which a buried quantum well gate performs
two functions: as a hole-collecting electrode and as a channel current modulation
gate.  The quantum well gate is made with a germanium layer deposited on a
silicon substrate. The active layers are of the order of 5 microns below the surface, permitting detection of minimum ionizing particles.  This technology is intended to achieve extremely small
pitch size to enable trigger-free operation without multiple hits
in a future linear collider, as well as simplified reconstruction of tracks with low
transverse momentum near the interaction point.

\bigskip
Thin film detectors  have the potential to be fully integrated,
while achieving large area coverage and low power consumption with low dead
material and low cost.  Thin flim transistor technology uses crystalline growth
techniques to layer materials, such that monolithic detectors may be fabricated
by combining layers of thin film detection material with layers of amplification
electronics using vertical integration. 

\bigskip
Lastly, a technology is under development in which a novel ultra-fast scintillating material employs a semiconductor stopping medium with embedded quantum dots.  The candidate material, demonstrating very high light yield and fast emission, is a GaAs matrix with InAs quantum dots. The first prototype detectors have been produced, and pending research goals include demonstration of detection
performance with minimum ionizing particles, corresponding to signals of about
4000 electron-hole pairs in a detector of 20 micron thickness.  A compatible electronics
solution must also be developed.  While the radiation tolerance of the device is
not yet known, generally quantum dot media are among the most radiation hard
semiconductor materials.  

\clearpage

\section{Introduction}

Research in particle tracking detectors for high energy physics application is underway with a goal of improving radiation hardness, achieving improved position, vertex, and timing resolution, simplifying integration, and optimizing power, cost, and material.  The five technologies described here approach these goals in complementary ways.

\section{Silicon Sensors in 3D Technology}

{\sl Boscardin, Dalla Betta, Hoeferkamp, Seidel, Sultan}

\bigskip
Silicon sensors of the 3D technology~\cite{DaVia} are employed in LHC experiments~\cite{Abbott,Ravera} to provide
radiation tolerant particle tracking at integrated fluences in the regime of $10^{16} n_{\rm eq}/{\rm cm}^2$. The
decoupling of the depletion depth from the sensor thickness allows operation at bias voltages
below breakdown despite very high integrated fluence, with significant savings on power
dissipation, and the small inter-electrode distance suppresses the effect of radiation-induced charge
trapping. The ATLAS IBL sensors, for example, are implemented in p-type with $230 \mu$m thickness
and column electrodes of diameter approximately $10 \mu$m, separated by approximately 62 microns. A slim edge
of 200 microns is employed. Designs for application to the HL-LHC, where innermost tracking
will be exposed over the course of 10 years to fluence $2.3 \times 10^{16} n_{\rm eq}/{\rm cm}^2$~\cite{Contardo}, are more aggressive still,
in anticipation of conditions in which the carrier lifetime will be reduced to 0.3 ns, corresponding
to a mean free path of 30 microns. Up to 200 interactions per 25 ns bunch crossing are expected
at the HL-LHC. Small-pitch 3D pixels ($25 \times 100~\mu {\rm m}^2$ or $50 \times 50~\mu{\rm m}^2$) have been developed to this
purpose, with inter-electrode distances of approximately 30 microns~\cite{DallaBetta} and a slim edge of 150 microns, and are
currently in the pre-production phase for the ATLAS ITk.

\bigskip
Plans~\cite{Riegler} for future facilities such as the FCC-hh anticipate a lifetime integrated luminosity
of 30 ab$^{-1}$, predicting integrated fluence at the innermost tracking volume approaching $10^{18}n_{\rm eq}/{\rm cm}^2$.
Estimates~\cite{Drasal} of the pileup conditions are on the order of 1000 events per crossing.
Continued development of silicon sensors of the 3D technology presents prospects both for
restoration of signal loss in high radiation environments, and for separation of pileup signals by precision
timing.  Measurements~\cite{Kramberger} carried out on $50 \times 50~\mu{\rm m}^2$ cell 3D sensors have shown signals with a
full width of 5 ns, and a rise time of 1.5 ns, with a timing resolution of 30 - 180 ps (depending on
the signal amplitude); this is a mode of operation comparable to that achieved by low gain
avalanche detectors --- but lacking gain --- with the advantage of higher radiation tolerance and
better fill factor. The standard column configuration of 3D has the disadvantages, however, that
the electric and weighting fields are non-uniform, leading to position dependence of the pulse
rise time; this is the limiting factor on the timing resolution. New geometries~\cite{Forcolin,Anderlini} involving p-type
trench electrodes spanning the entire length of the detector, separated by lines of segmented
n-type electrodes for readout, promise improved uniformity and better timing resolution combined
with further increased radiation tolerance. Nevertheless, at this time, trenched electrodes cause
higher capacitance and introduce larger dead volumes within the substrate. Device optimizations,
especially in terms of geometrical efficiency, remain to be carried out. In addition, this problem
can be tackled at the system level by tilting the sensor plane with respect to the particle direction,
so that a larger fraction of the charge is generated within the depleted volume, and using multiple
planes of sensors with an offset between the electrodes, so all tracks would traverse several planes
without crossing the electrodes~\cite{Mendicino}.

\bigskip
3D columnar pixels with internal gain~\cite{Feasibility}-\cite{Indication} offer an alternative approach to signal
restoration at high fluence. When implemented with very small inter-electrode separation, approximately 15
microns or less, these devices can achieve controlled charge multiplication at voltages on the order
of 100 V, both before and after irradiation. Moderate gain values can be achieved, sufficient to
compensate the loss of charge signal due to irradiation of these thin (approximately $100~\mu$m) devices. Design
optimization continues with a goal of achieving uniform gain throughout the cell active volume,
also benefiting from the wider operating range that is possible due to increasing the breakdown
voltage.

\bigskip
The goal of this research is to advance one or two 3D technologies in silicon for tracking particles,
able to operate with adequate signal-to-noise ratio at fluences approaching $10^{18}$ n$_{\rm eq}/{\rm cm}^2$, and timing
resolution on the order of 10 ps. Planned research activities include TCAD simulations, process
optimization and fabrication of several generations of prototypes, and thorough characterization
of the prototypes before and after irradiation to extreme fluences.

\section{3D Diamond Detectors}

{\sl Kagan, Trischuk}

\bigskip
By 2028, experiments operating at the HL-LHC must be prepared for an instantaneous luminosity of
$7.5 \times 10^{34}/{\rm cm}^2/{\rm s}$ and charge particle fluxes of GHz/cm$^2$.  After these doses, all detector materials will
be trap-limited, with the average drift distance a free charge carrier travels before it gets trapped being
below $50~\mu$m~\cite{Tsung}.  3D sensors reduce the drift distance the charge carriers must travel to reach an electrode to much less
than the sensor thickness. This is
particularly beneficial in detectors with a limited distance free charge carriers travel, such as
trap-dominated sensor materials like heavily irradiated silicon and pCVD diamond, where the observed
signal size is related to the mean free path divided by the drift distance. Under these circumstances
one gains radiation tolerance (larger signals) by keeping the drift distance less than the mean free
path. With the 3D geometrical structure, charge carriers drift inside the bulk parallel to the surface
over a typical drift distance of $25-100~\mu$m instead of perpendicular to the surface over a distance of
$250-500~\mu$m.

\bigskip
The RD42 collaboration has studied novel 3D detector designs in diamond, to extend the radiation
tolerance of diamond to fluences greater than $10^{17}$ hadrons/cm$^2$, exceeding the HL-LHC doses. The
detector design places column-like electrodes inside the detector material using a 130 fs laser with a
wavelength of 800 nm. After focusing to a $2~\mu$m spot, the laser has the energy density to convert
diamond into an electrically resistive mixture of different carbon phases~\cite{Pimenov}. A Spatial Light
Modulator (SLM)~\cite{Sun} is used to correct spherical aberrations during fabrication. This helps to
achieve in $50~\mu{\rm m} \times 50~\mu{\rm m}$ cells a high column yield of $\ge 99.8$\%, a small
column diameter of $2.6~\mu$m, and a resistivity of the columns of the order of $0.1-1~\Omega$cm.
In this detector geometry, the drift
distance an electron-hole pair must travel to reach an electrode can be reduced below the mean free
path of an irradiated sensor without reducing the number of electron-hole pairs created. In a detector
with $25~\mu{\rm m} \times 25~\mu{\rm m}$ cells, the maximum drift distance for charge carriers that go into the saddle
point region is $25~\mu$m, and $17.5~\mu$m for charge carriers that avoid the saddle point.
The goal of this research project is to create a detector that is essentially immune to radiation
doses at the level of $10^{17}$ hadrons/cm$^2$. Initial tests have shown that after $3.5 \times 10^{15}$ n/cm$^2$, the 3D
geometry with $50~\mu{\rm m} \times 50~\mu{\rm m}$ cells has better than three times less charge loss than a planar diamond detector after
normalizing both unirradiated devices to a relative charge of 1. Furthermore the charge in the
unirradiated 3D device is twice as large as that in the planar device. Thus, in addition to having twice
the charge, the 3D device also has better than three times less damage, due to the shorter drift distance. In order to
achieve the $10^{17}$ hadrons/cm$^2$ goal, completion of the design of 3D diamond devices with
$25~\mu{\rm m} \times 25~\mu{\rm m}$ cells and testing of these devices after irradiation with $10^{17}$ hadrons/cm$^2$ is proposed.

\section{Submicron Pixels with a Quantum Well for Vertexing}

{\sl Fourches, Renard, Barbier}

\bigskip
Development of a submicron position sensitive vertex detector for the future linear collider experiments is proposed.
Although improved relative to
their predecessors, the present vertexing pixel detectors at the Large Hadron Collider suffer from low position resolution. The objective of vertex detectors is to enable the accurate secondary vertex determination that is crucial for b-tagging~\cite{Nagai,Pasquali}
in the case of high transverse momentum ($p_{\rm T}$) events. The heavy quark events are characterized by a relatively high lifetime that leads to
a secondary vertex distinct from the interaction point~\cite{Tanabe}. For accurate track reconstruction, it is necessary to improve point-to-point resolution
well below the 5 micrometer limit. In the framework of ILD, development of a pixel detector based on the DoTPiX structure is proposed.

\bigskip
Accurate track reconstruction with a vertex detector is possible using a small pitch detector, which in the case of the ILC can reduce the
multiplicity (in which a pixel is hit several times). This is crucial for the ILD where the readout of the detector is made only after several bunches.
With a track fit, displaced secondary vertices can be evaluated, using an impact parameter technique to select the right track, and the analysis of the full decay
of the particle can be done, using all necessary jets. In addition, isolated tracks can be tagged in order to reduce fake events.
The vertex detectors implemented in the LHC experiment are based on a hybrid design.
The high particle rate at the LHC induces a large dose in the detector where non-ionizing energy loss damages
the detector material and the electronic readout. Special techniques have been used to circumvent these effects with the use of hardened processes~\cite{Dentan}
and adequately doped silicon pixel structures~\cite{Terzo}.
To accommodate the LHC beam crossing time, detectors use a triggered readout involving a fast readout chip (ROC).
The on-pixel electronics has to be elaborate to collect the information of all the pixel hits' output when triggered.
For the technologies available from the late 1990's to the early 2000's, this requirement excluded small pitch pixel detectors.
Even with pitch of tens of microns, the number of channels (pixels) is of the order of tens of millions in the inner vertex detectors.

\bigskip
The constraints are different for the ILC where more precise reconstruction is the objective. The advantage of vertex detectors with much improved
resolution will be good secondary vertex reconstruction with an accuracy of 0.5 micrometers (or in time at the speed of light, of 1.6 fs).
This cannot be matched by a timing procedure, which can only estimate the position of the interaction point in the beam-crossing zone.
This zone will be reduced at the ILC compared with the LHC. Additionally, short-lived particles can be tagged at this stage.
Significant features of this proposal include the following.

\begin{itemize}
\item The detectors close to the primary interaction point can detect low-mass charged particles that can escape the tracker due to
the effect of the magnetic field on low-mass particles~\cite{Griso}.
Tagging of such particles can be established with a good vertex detector~\cite{Suehara}.
These particles can produce disappearing tracks. The energy of these long-lived particles cannot be determined easily as they escape calorimetry.
The only possibility, besides using time-of-flight, is to add extra layers to the vertex detector to match the trajectory.

\item The operation of a vertex detector in a trigger-free mode means that many bunch crossings will be combined (this is pile-up) before being output and reset.
This makes the use of very small pixels necessary to avoid multiple hits in a single pixel. The pitch has to be reduced to match these requirements,
and only a fully monolithic pixel can be used for this purpose.

\item The reconstruction of tracks with relatively low $p_{\rm T}$ near the interaction point will be easier with a pixel detector with large enough
aspect (height/width) ratio. A small pitch (less than 1 micrometer) with a height up to of 10 micrometers (the sensitive zone) opens up such a possibility.
\end{itemize}

\begin{figure}[htbp]
  \centering
      \includegraphics[width=1.0\linewidth]{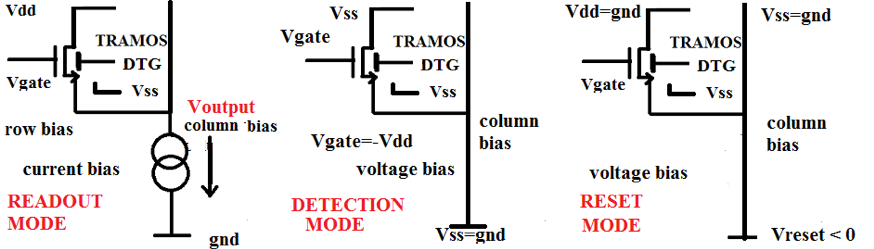}
	\caption{The operational principle of the DoTPiX structure within a pixel array (row and column); the array readout is similar to those of CMOS sensors, with detection, readout, and reset modes.  The end of column is connected to a preamplifier, for digital or hit/no hit readout mode. Power dissipation occurs only during readout, due to the biasing scheme. In detection mode, $V_{\rm gate} < V_{\rm drain}$  and $V_{\rm source}$, to collect holes in the buried gate.}
	\label{Fourches-figure1}
\end{figure}

\begin{figure}[htbp]
  \centering
      \includegraphics[width=1.0\linewidth]{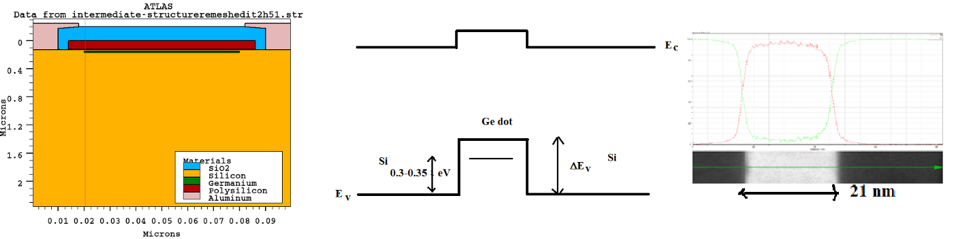}
	\caption{For the DoTPiX project: (left) the TCAD simulation structure; (center) Ge hole quantum well, and (right) results of the processing (on a full wafer), the deposition of a thin Ge layer.
	 This results from electron microscopy, STEM Energy Dispersive X Spectrometry (STEM-EDX).  The Ge concentration reaches 95\% in the 21 nm buried layer. The wafer prepared this way should be CMOS compatible with attention to the thermal budget of the process.}
	\label{Fourches-figure2}
\end{figure}

\bigskip
A pixel design has been proposed~\cite{Fourches1,Fourches2}. The necessary simulations have been made to assess the functionality of the proposed device. The next step is to find
out what is the best process to obtain the functionality and to reach some specifications.  The principle of this pixel architecture (named DoTPiX) is the single n-channel MOS transistor, in which a buried quantum well gate performs two functions (Fig.~\ref{Fourches-figure1}): as a hole-collecting electrode and as a channel current modulation gate. Extensive simulations were made with this pixel architecture~\cite{Fourches1}; to assess its functionality, the buried quantum well gate is made with a Ge layer deposited on a silicon substrate. We are currently developing the basic structure with UHV/CVD techniques. The proposed structure is in its fabrication phase to obtain a test vehicle. The active layers where electron-hole pairs are created is of the order of $5~\mu$m below the surface, enabling detection of minimum ionizing particles. The processed wafers (Fig.~\ref{Fourches-figure2}) should be compatible with advanced CMOS nodes including SOI (FDSOI) and nanowire devices. The surface roughness on the processed wafers (as measured by AFM) is low enough but should be slightly improved.  We have established a working group with different institutes for this project.

\bigskip

\clearpage

\section{Thin Film Detectors }

{\sl Kim, Metcalfe, Sumant}

\bigskip
Nanoscience technologies are developing new cutting edge materials and devices for a wide range of applications.
HEP can take advantage of the many advances by looking toward thin film fabrication
techniques to implement a new type of particle detector. Thin Film (TF) Detectors have the potential to be fully
integrated, large area, low power, with low dead material, and low cost. The present goal is to investigate potential
research paths using thin film technologies and to identify and characterize the performance benefits for
future particle experiments.

\bigskip
A new detector technology is proposed based on thin films that is aimed at dramatically
improving the precision of particle detectors by greatly reducing the mass of the detector~\cite{Metcalfe6}. Cleaner signatures
of the particles from the primary collision will be obtained by reducing those particles' interactions
with dead material, which will improve reconstruction efficiencies and resolutions.
Thin Film technologies could potentially replace the entire detector including all the services. If a thin
film detector could be printed in large areas (square meters), it is estimated that the cost would be reduced to less than 1\%
of the current cost. If the nuclear interaction length can be decreased by a factor of 10, then the track
reconstruction efficiencies would reach 99\% and enable a host of new measurements and searches.

\bigskip
Thin Film technology presents one possible solution to achieve
these performance milestones. TF transistors (TFT’s) were first conceived in the 1960’s by Paul Weimer~\cite{Weimer}.
By the 21st century, fabrication technologies had improved enough to make it competitive
with existing technologies. TFT’s are the basis of technologies such as Liquid Crystal Display (LCD)
screens, solar cells, and light emitting diodes. It is a rapidly growing technology area with a large market
base and has corresponding investment in large scale fabrication and industrialization. Ultimately, the
broader interest of these technologies enables HEP to leverage the investments in commercialization as well
as the R \& D into materials, tools, and techniques.

\bigskip
Some of the advantages of TF’s are optical transparency, mechanical flexibility, high spatial resolution,
large area coverage, and low cost relative to traditional silicon-based semiconductor technology. TF technology
uses crystalline growth techniques to layer materials. Monolithic sensors can be fabricated using layers
of thin film materials for particle detection with layers for amplification electronics. The advantages of a
detector made with this type of technology include single piece large area devices (on the order of a few
square meters), high resolution ($< 10~\mu$m), low cost (100 times less than that of Si-CMOS), low mass, and high curvature for a
cylindrical, edgeless design~\cite{Gnade}-\cite{Street}.

\bigskip
Fabrication processes such as chemical bath deposition and close-space sublimation on a substrate material
can produce thin films with a high degree of precision. Here, the crystalline structure is grown in
layers, avoiding drilling and etching techniques standard in traditional silicon fabrication; consequently, TF
processing is much less expensive.

\begin{figure}
    \centering
    \includegraphics[width=0.7\linewidth]{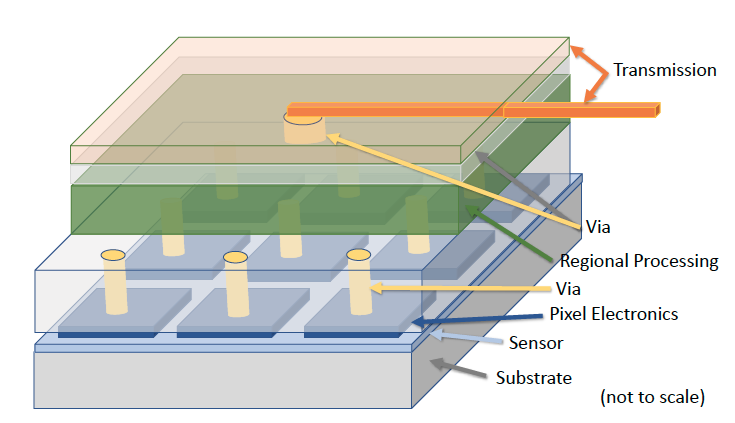}
    \caption{The conceptual stack-up of a monolithic thin film detector that
      incorporates suppot structure, sensor, pixel electronics, regional
      processing electronics, and power/data transmission.  Potentially, no
      other structures would be needed in the detector volume.}
    \label{Metcalfe-figure}
\end{figure}

\bigskip
Thin film electronics can be vertically integrated with a thin film sensor if the fabrication
techniques are compatible. This would allow vertical integration of sensor and pixel electronics. Further
vertical integration using through-vias would enable signals to pass from one layer to the next, thus enabling
several levels of electronic processing. Typical front-end ASIC functions could be integrated into the
monolithic structure as well as higher-end processing to perform functions such as data aggregation and
region-of-interest processing. Such processing would reduce the number of transmission lines integrated
into a top layer and further reduce the material inside the detector volume.
Figure~\ref{Metcalfe-figure} shows a potential vertical stack-up.

\bigskip

The transistor is the most basic unit that determines the power consumption in electronics. Complementary Metal Oxide Semiconductor
(CMOS) is a low power technology and is the current mainstay for most of commercial
electronics. There are, however, many types of transistor technologies that can outperform CMOS.
Silicon Germanium (SiGe) Heterojunction Bipolar Transistors (HBT’s) are another class of transistor that
typically boast faster speeds and lower power consumption~\cite{Cressler}-\cite{Metcalfe5}. Fin-Field Effect Transistors (FinFET’s) are
being pursued as the next ultra-low power technology and are manufactured by companies such as IBM and
Motorola. However, the most transformational transistor is the Thin Film Transistor (TFT), which is breaking
records in terms of size and power. All of these technologies have the potential for reducing the power (and
the copper in the transmission lines) over the current technologies.

\bigskip
Thin Film Detectors have the potential to replace a wide range of detector types
from tracking to calorimetry. The present goals are to identify key areas of research within
Thin Film technologies, quantify the key requirements from different types of experiments, and evaluate the
potential physics impact.

\clearpage

\section{Scintillating Quantum Dots in GaAs for Charged Particle Detection}

{\sl Mahajan, Minns, Tokranov, Yakimov, Oktyabrsky, Gingu, Murat, Hedges}

\bigskip
Future collider experiments will require particle detectors with timing resolution better than 10 ps; this is beyond the limits of existing technologies \cite{BRN2019}. One possible avenue for innovation of charged-particle tracking relies on novel ultra-fast scintillating material utilizing semiconductor stopping media with embedded quantum dots \cite{Oktyabrsky2016, Dropiewski2020, DROPIEWSKI2020161472}.

\bigskip
Fabrication of such a detector requires a scintillator with unique properties: very high light yield and a fast emission time. We have identified a candidate sensor material based on self-assembled InAs quantum dots (QDs) embedded into a GaAs matrix \cite{Oktyabrsky2016}. QDs are known to be excellent light emitters with close to 100\% efficiency and emission times on the order of hundreds of picoseconds. To make a scintillator, however, one needs to embed QDs into a dense medium that is transparent to the QD photon emission. GaAs fulfills this requirement. An ideal sensor consists of two physically integrated systems:

\bigskip
\begin{enumerate}
\item{The scintillator: A charged particle travels through the InAs QD/GaAs scintillator and produces electron-hole pairs in the GaAs matrix ($2.4 \times{} 10^5$ pairs/MeV). The carriers are rapidly captured (within 2-5 ps) in the positively charged QDs due to high electron mobility of up to 8500 $\rm{cm}^2 / \rm{Vs}$ at room temperature. The infrared emission (1.1 eV photons) is red-shifted more than 300 mV from the bandgap of the GaAs matrix, resulting in low self-absorption ($\sim{}1 \rm{cm}^{-1}$) \cite{Dropiewski2020}.}

\item{The photodetectors: As the refraction index of GaAs is high (approximately 3.4), only $\sim{}2\%$ of the emitted light exits the
scintillator through one planar interface with air, and the rest gets reflected and travels inside the scintillator. Thus for efficient detection, the photodetector (PD) must be physically integrated with the scintillator. The proposed design has a matrix of InGaAs photodiodes fabricated directly on the surface of the scintillation matrix. The photodiode thickness is of the order of 1--2 microns, leading to efficient absorption of the QD emission. Photodiodes fully cover the scintillator area, resulting in uniform and efficient collection of the emitted light with close to unity fill factor.}
\end{enumerate}

\begin{figure}[htbp]
  \centering
\includegraphics[width=0.6\linewidth]{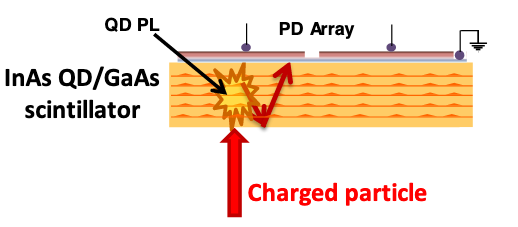}
\caption{A schematic drawing of the proposed tracking sensor. A charged particle enters the GaAs scintillator, producing electron-hole pairs. The electrons are then quickly trapped by the positively charged InAs quantum dots (QDs). The QDs undergo photoluminescence (PL) and emit photons that travel through the medium. The emitted photons are collected by a photodiode (PD) array.}
\label{fig:qds-schem}
\end{figure}

\begin{figure}[htbp]
    \centering
    \includegraphics[width=0.7\linewidth]{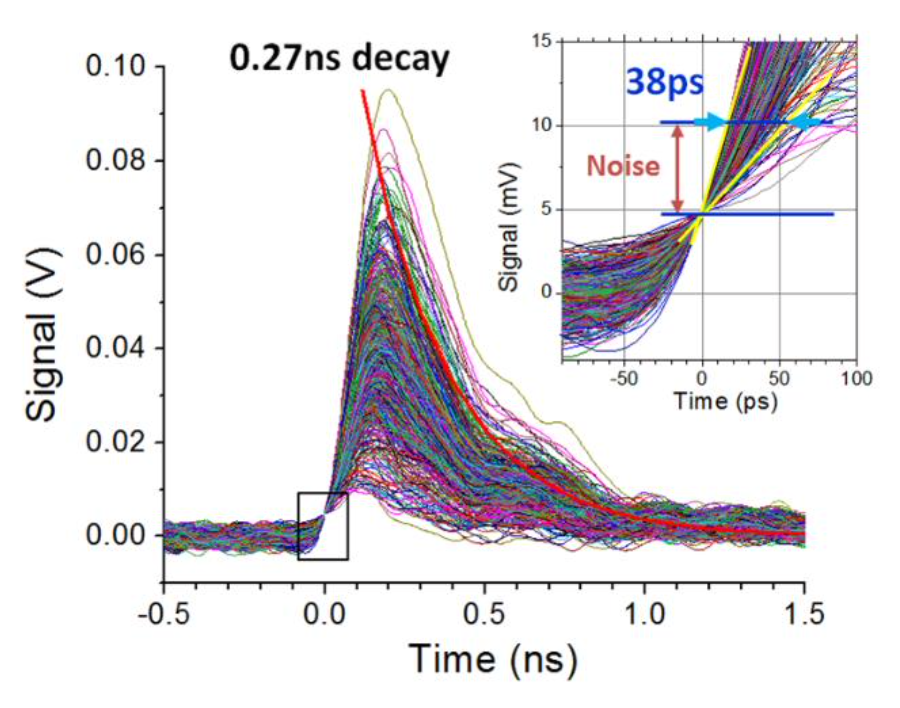}
    \caption{Scope traces of events from $\rm{Am}^{241}$ events. Recorded pulses showing 100 ps rise time, 270 ps decay time, and 38 ps time resolution with average collected charge of $1.5\times10^5$ electrons. Adapted from Ref. \cite{Mahajan2021}.}
    \label{fig:qds-timingres}
\end{figure}

\begin{figure}
    \centering
    \includegraphics[width=0.7\linewidth]{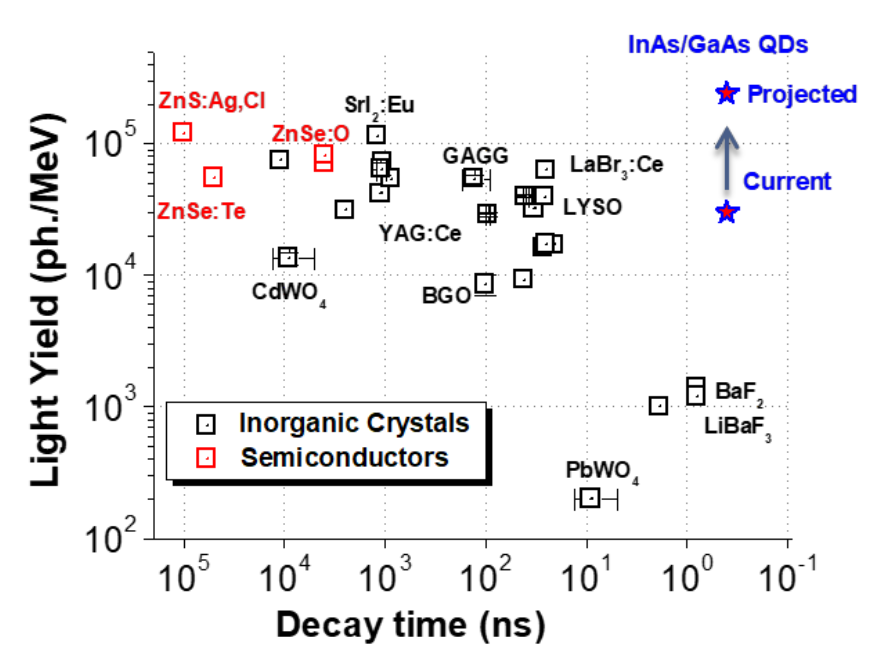}
    \caption{Light yield in photons/MeV of deposited energy versus decay time for various known scintillators. Faster performance (decreasing decay-time) extends rightward along the horizontal axis. Adapted from Ref. \cite{Mahajan2021}.}
    \label{fig:qds-performance}
\end{figure}

A schematic drawing showing this system is shown in Fig.~\ref{fig:qds-schem}.
The first prototype detectors have been produced at SUNY Polytechnic Institute at Albany as thin wafers of $\sim20$ micron thickness, each with a single integrated small area PD. The measurements of single-channel performance with $\alpha$ particles have been published~\cite{Dropiewski2020, DROPIEWSKI2020161472, Minns2021, Mahajan2021}. Using $\mathrm{Am}^{241} ~\alpha$-sources (5.5 MeV) and fast preamplifiers, we have measured a fast-decay constant of 270 ps and a 38 ps time resolution at room temperature without bias voltage applied to the PD. This performance, shown in Fig.~\ref{fig:qds-timingres}, is currently limited by circuit noise and bandwidth. We measure $1.7 \times 10^4$ detected electrons per 1 MeV of deposited energy with this system. Alternatively, a slower low-noise preamplifier demonstrated $5.1 \times 10^4$ electrons/MeV of incident energy with longer ($\sim6$ ns) pulses. We expect performance to be near that of the theoretical optimum of $2\times10^5$ electrons/MeV. A comparison of the performance of this detector with that of other scintillators is shown in Fig.~\ref{fig:qds-performance}. The demonstrated performance is, at the time of writing, the fastest and highest light yield of all known scintillators.

\bigskip

Significant exploratory research and development is required to accurately assess expected performance of these detectors in future high-energy physics applications. First, we must demonstrate detection performance with minimum ionizing particles, corresponding to expected signals of about 4000 electron-hole pairs in a single detector of $20~\mu$m thickness. Given that the measurements with $\alpha$-particles are noise-limited, we also expect to encounter significant challenges developing a suitable electronics solution for optimal energy and timing performance for MIP detection. Furthermore, the radiation tolerance of this type of custom epitaxially-grown detector is not known, although the QD medium itself is among the most radiation hard semiconductor materials \cite{Oktybrsky2005}. We will ultimately need to assess the performance of these detectors in the high-radiation environments expected in future high-energy physics experiments.
  
\section{Conclusion}

Five contemporary technologies are under development for applications at future
high energy physics experiments.  Collaborators interested in joining any of these efforts are welcome.








\end{document}